\documentclass{sig-alternate}
\usepackage{url}
\usepackage{xspace,colortbl}
\usepackage[markup=nocolor]{changes}
\usepackage[english]{babel}
\usepackage{multirow}
\usepackage{subfigure}
\usepackage{graphicx}
\usepackage{amssymb}
\usepackage{fmtcount}
\usepackage{amsfonts}
\usepackage{xspace}
\usepackage{amsmath}
\usepackage{multirow}
\usepackage[mathscr]{eucal}
\usepackage{colortbl}
\usepackage{bm}
\usepackage[nospace]{cite}
\makeatletter
\newif\if@restonecol
\makeatother
\makeatletter
\def\@copyrightspace{\relax}
\makeatother

\usepackage[lined,boxed,vlined,ruled]{algorithm2e}

\long\def\comment#1{}

\begin{document}

\clubpenalty=10000
\widowpenalty = 10000

\tolerance=1000


\title{The Expected Optimal Labeling Order Problem for Crowdsourced Joins and Entity Resolution}

\newtheorem{theorem}{Theorem}
\newtheorem{example}{Example}
\newtheorem{definition}{Definition}
\newtheorem{proposition}{Proposition}
\newtheorem{lemma}{Lemma}
\newtheorem{corollary}{Corollary}
\newcommand{\crowdert}{$\textsf{CrowdEST}$\xspace}

\newcommand{\dataset}{data set\xspace}
\newcommand{\datasets}{data sets\xspace}
\newcommand{\rp}{record pair\xspace}
\newcommand{\rps}{record pairs\xspace}
\newcommand{\order}{\omega\xspace}
\newcommand{\orderlabel}{\omega\ell\xspace}
\newcommand{\labelset}{\mathcal{L}\xspace}
\newcommand{\publish}{\mathcal{P}\xspace}
\newcommand{\clustergraph}{\textsc{ClusterGraph}\xspace}
\newcommand{\clustergraphs}{\textsc{ClusterGraphs}\xspace}
\newcommand{\cluster}[1]{\textsf{cluster}(#1)\xspace}
\newcommand{\clusterp}[1]{\textsf{cluster'}(#1)\xspace}
\newcommand{\cn}{\mathcal{C}\xspace}
\newcommand{\ecn}{\mathrm{E}\xspace}
\newcommand{\nphard}{$NP$-$hard$\xspace}
\newcommand{\paper}{{\textsf{Paper}}\xspace}
\newcommand{\cora}{{\textsf{Cora}}\xspace}
\newcommand{\transitive}{{\textsf{Transitive}}\xspace}
\newcommand{\nontransitive}{{\textsf{Non-Transitive}}\xspace}
\newcommand{\optimalorder}{{\textsf{Optimal Order}}\xspace}
\newcommand{\parallelno}{{\textsf{Parallel}}\xspace}
\newcommand{\nonparallel}{{\textsf{Non-Parallel}}\xspace}
\newcommand{\parallelid}{{\textsf{Parallel(ID)}}\xspace}
\newcommand{\parallelnf}{{\textsf{Parallel(ID+NF)}}\xspace}
\newcommand{\expectoptimalorder}{{\textsf{Expect Order}}\xspace}
\newcommand{\randomorder}{{\textsf{Random Order}}\xspace}
\newcommand{\worstorder}{{\textsf{Worst Order}}\xspace}
\newcommand{\product}{{\textsf{Product}}\xspace}
\newcommand{\abtbuy}{{\textsf{Abt-Buy}}\xspace}
\newcommand{\tp}{{\textsf{tp}}\xspace}
\newcommand{\precision}{{\textsf{precison}}\xspace}
\newcommand{\recall}{{\textsf{recall}}\xspace}
\newcommand{\fp}{{\textsf{fp}}\xspace}
\newcommand{\fn}{{\textsf{fn}}\xspace}
\newcommand{\prob}{\mathrm{P}\xspace}
\newcommand{\addtext}[1]{{{\textcolor{blue}{ #1}}\xspace}}
\pagestyle{plain}

\numberofauthors{1}
\author{\alignauthor Jiannan Wang{$\,^{\#}$},~~ Guoliang Li{$\,^{\#}$},~~ Tim Kraska{$\,^\dag$},~~ Michael J. Franklin{$\,^\ddag$},~~ Jianhua Feng{$\,^{\#}$} \\
\vspace{.2em}\affaddr{$^{\#}$Department of Computer Science,  Tsinghua University, ~~ $^\dag$Brown University, ~~ $^\ddag$AMPLab, UC Berkeley} \\
\vspace{.1em}\email{wjn08@mails.tsinghua.edu.cn,~~ligl@tsinghua.edu.cn,~~tim\_kraska@brown.edu}\\
\email{franklin@cs.berkeley.edu,~~~~fengjh@tsinghua.edu.cn}
}

\maketitle
\thispagestyle{plain}


In the SIGMOD 2013 conference, we published a paper~\cite{DBLP:conf/sigmod/WangLKFF13} extending our earlier work on crowdsourced entity resolution to improve crowdsourced join processing by exploiting transitive relationships. The VLDB 2014 conference has a paper~\cite{DBLP:journals/pvldb/NorasesBD14} that follows up on our previous work, which points out and corrects a mistake we made in our SIGMOD paper. Specifically, in Section 4.2 of our SIGMOD paper, we defined the ``Expected Optimal Labeling Order" (EOLO) problem, and proposed an algorithm for solving it.  We incorrectly claimed that our algorithm is optimal. In their paper, Vesdapunt et al. show that the problem is actually NP-Hard, and based on that observation, propose a new algorithm to solve it.

In this note, we would like to put the Vesdapunt et al. results in context, something we believe that their paper does not adequately do. 

\section{Contributions of the SIGMOD 2013 Paper}
\begin{figure}[!h]
\centering
  \includegraphics[scale=0.95]{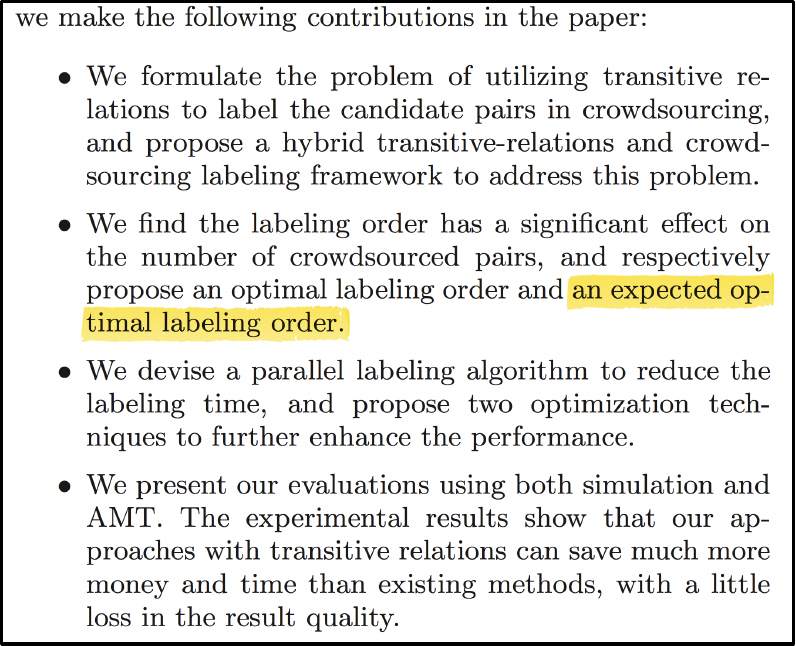}\\\vspace{-1em}
  \caption{Contributions of the SIGMOD 2013 paper.}\label{fig:contributions}\vspace{-1em}
\end{figure}

The main contributions of our SIGMOD 2013 paper were to identify the importance of exploiting transitivity to reduce the cost and improve the performance of crowdsourced join processing and to present a new framework for implementing this technique. The issue addressed by Vesdapunt et al. concerns only one aspect of our claimed contributions (shown in Figure~\ref{fig:contributions}) namely, one sub-point (highlighted) of the second (of four) contributions listed in the SIGMOD paper. All of the other contributions are unaffected. 

In particular, the NP-hardness of the EOLO problem does not affect either the correctness of the framework or the correctness of our experimental results. It only means that the efficiency of one of the components in the framework we proposed can be improved, which the authors of the VLDB paper went ahead and did.

\section{The Bug in the EOLO optimality proof}
In our proof of optimality for our solution to the EOLO problem, we included several situations that could not occur in practice. Below, we use an example from Vesdapunt et al., to illustrate the problem. 



Suppose we have three pairs: (a, b), (a, c), and (b, c). We want to label for each pair whether it refers to the same entity or not. There are two ways to label these pairs. One is to ask the crowd a question such as ``whether a pair (e.g., a and b) refers to the same entity". The other way is to use transitive relations to deduce the pair's label. For example, if we know ``$a = b$" and ``$b = c$", then we can deduce ``a = c" without asking the crowd to label it. Similarly, if we know ``$a = b$" and ``$b \neq c$", then we can deduce ``$a \neq c$" using transitive relations as well. For a given pair, if its label is obtained from the crowd, we call it as ``crowdsourced pair"; if its label is deduced based on transitive relations, we call it as ``deduced pair". 

Given three pairs: (a, b), (a, c), and (b, c). Suppose each of the pairs has a probability of 0.5 to refer to the same entity. Consider a labeling strategy, which labels the pairs one by one in the order of (a, b) -> (a, c) -> (b, c), and asks the crowd to label a pair iff. its label cannot be deduced from transitive relations. The question is: ``What is the expected number of crowdsourced pairs for the labeling strategy?"

\begin{figure}[htbp]
\centering
  \includegraphics[scale=0.9]{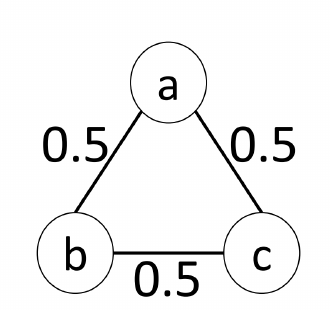}\\ \vspace{-.5em}
  \caption{The probabilities of (a, b), (b, c) and (a,~c).}\label{fig:pairs} 
\end{figure}

To solve this problem, we first compute the probability of each pair being a crowdsourced pair, and then sum up their probabilities. 
\begin{itemize}
\item For the first pair ``(a, b)", as there is no labeled pair before it, we have to ask the crowd to label it, thus the probability of (a, b) being a crowdsourced pair is equal to 1. 
\item For the second ``(a, c)", since its label cannot be deduced from the first pair (a, b), we have to ask the crowd to label it, thus the probability of (a, c) being a crowdsourced pair is equal to 1 as well.
\item For the third pair ``(b, c)", it needs to be crowdsourced only when both of ``$a\neq b$" and ``$b\neq c$" hold. Since the probability of ``a = b" is 0.5 and the probability of ``b = c" is 0.5, the probability of the event that both of ``$a\neq b$" and ``$b\neq c$" hold is (1-0.5)*(1-0.5) = 0.25. Thus, we calculate that the probability of (b, c) being a crowdsourced pair is 0.25.
\end{itemize}

By summing up the three probabilities, the expected number of crowdsourced pairs was computed as 1 + 1 + 0.25 = 2.25. This calculation turns out to be incorrect. The correct answer, as pointed out by Vesdapunt et al., is 2.4.

\begin{figure}[tbp]
\centering
  \includegraphics[scale=0.4]{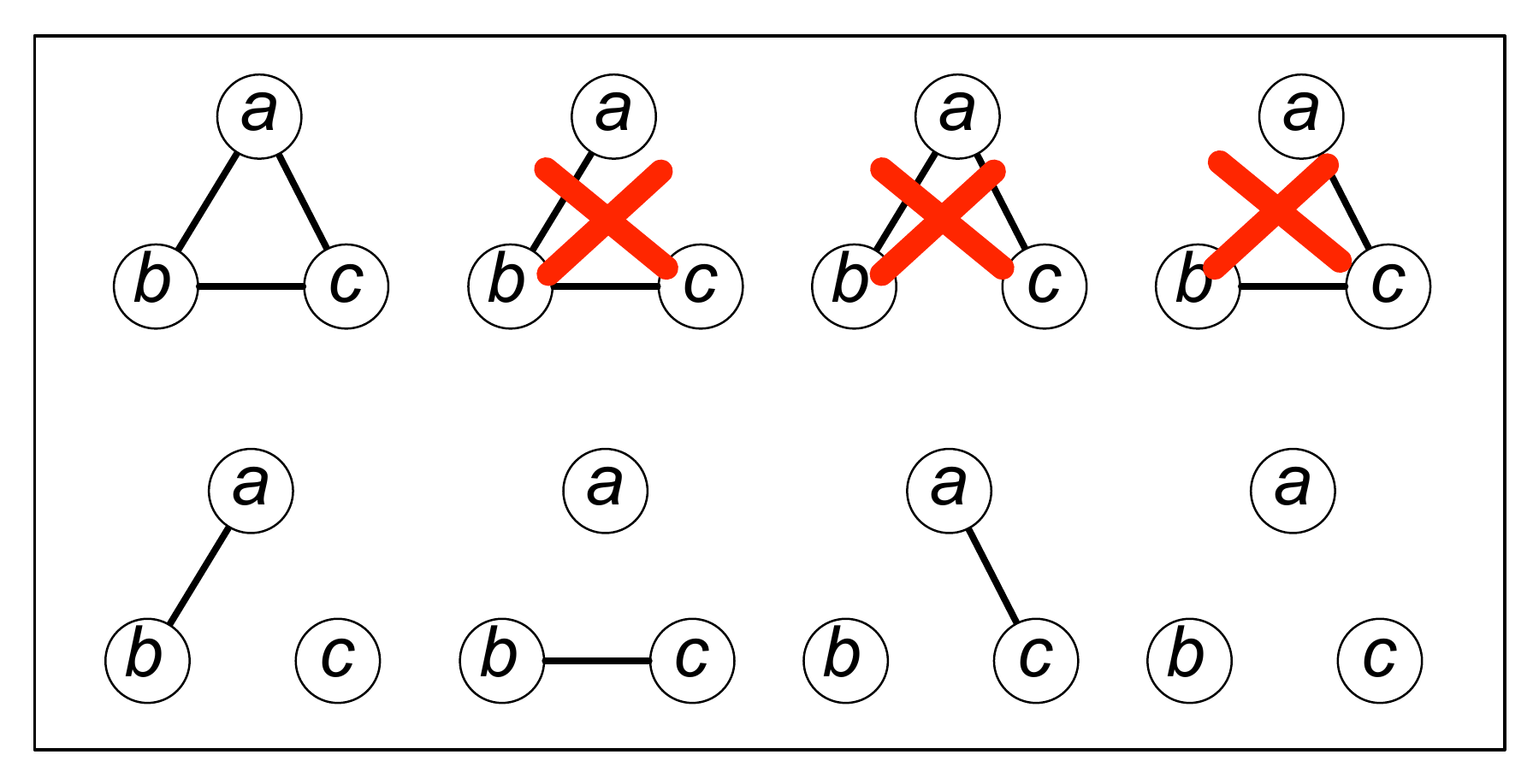}\\
  \caption{Five possible cases of the labels of (a, b), (a, c), and (b, c).}\label{fig:worlds}\vspace{-1em}
\end{figure}

Our error was in the computation of the probability for the third pair (b, c). Consider the five possible cases of the labels of (a, b), (a, c), and (b, c) in Figure~\ref{fig:worlds}.  In the figure, an edge between a pair of nodes means that the two nodes represent the same entity. Note that the 2nd, 3rd, and 4th graphs are impossible since they violate the transitivity assumption. For the five possible cases, we can compute that each case has a probability of 1/5 = 0.2.  Since the third pair ``(b, c)" needs to be crowdsourced only when both of ``$a\neq b$" and ``$b\neq c$" hold, i.e., the 6th or the 8th graphs, the probability of (b, c) being a crowdsourced pair should be 0.2+0.2 = 0.4 instead of 0.25 as computed above. 

The incorrect calculation above led us to an incorrect proof of optimality for our solution to the EOLO problem and Vesdapunt et al. rightly show that under the transitivity assumptions we made, the problem is in fact NP-Hard. Furthermore, based on this insight they propose a new algorithm for this aspect of our framework.

\section{Algorithmic Comparisons}
In the VLDB paper, the authors first compared their algorithm with our algorithm on three real data sets. Their experimental results showed that on one dataset, their algorithm is preferable; on a second dataset, our algorithm performs better; on the third dataset, a simple random algorithm performs the best (Figure 12 in their paper). Next, the authors constructed a worst case (for our algorithm), and then showed that in this case their algorithm performed much better than ours (Figure 13). Unfortunately, in the introduction to their paper, Vesdapunt et al. make the claim that the performance of their algorithm is {\bf{an order of magnitude}} better than ours {\bf{in practice}}. We do not believe that their experimental results support this claim. In fact, that result was obtained only on an artificial scenario specifically created to defeat our algorithm.

\section{Updating our SIGMOD 2013 Paper}
Based on the analysis of Vesdapunt et al., we revised our SIGMOD 2013 to remove the claim of optimality for our EOLO solution and placed the revised paper on arXiv (\url{http://tiny.cc/revised}). 
The following are the changes that we made to the paper: 
\begin{itemize}
\item We corrected the example of computing the expected optimal labeling order in Section 4.2.
\item We cited the new VLDB paper in Section 4.2 to clarify that the EOLO problem is NP-hard.
\item We removed the claim that ``our algorithm can identify the expected optimal labeling order" from the paper. 
\end{itemize}

\subsection{Summary}
We appreciate that Vesdapunt et al. discovered flaw in an important aspect of our SIGMOD 2013 paper and were able to publish a full VLDB paper to address this issue.  We do feel strongly, however, that the introduction section of that paper overstates the relative benefits of their proposed algorithm for the EOLO problem relative to our original algorithm in practice.   That being said, it is clear that the research topic of crowdsourced query processing is gaining increasing attention and that there are a wide range of open challenges to be researched. For interested readers, we collected a list of papers published recently in this topic and put them on this link (\url{http://tiny.cc/crowdpaper}). 


{
\bibliographystyle{abbrv}
\bibliography{crowdsourcing}
}
\end{document}